\documentclass[aps,prb,floatfix,epsfig,twocolumn,showpacs,preprintnumbers]{revtex4}
\usepackage{amsmath,amssymb,graphicx,bm,epsfig}
\usepackage{color}

\newcommand{\vk}{{\mathbf{k}}}
\newcommand{\vK}{{\mathbf{K}}}
\newcommand{\vq}{{\mathbf{q}}}

\newcommand{\Tr}{\mathrm{Tr}}
\renewcommand{\a}{\alpha}
\renewcommand{\b}{\beta}
\newcommand{\g}{\gamma}
\renewcommand{\d}{\delta}

\begin{document}

\title{Quantum Monte Carlo Impurity Solver for Cluster DMFT and Electronic
  Structure Calculations in Adjustable Base}
\author{Kristjan Haule}
\affiliation{Department of Physics, Rutgers University, Piscataway, NJ 08854, USA}
\date{\today}

\begin{abstract}


We generalized the recently introduced new impurity solver
\cite{Werner_1} based on the diagrammatic expansion around the atomic
limit and Quantum Monte Carlo summation of the diagrams. We present
generalization to the cluster of impurities, which is at the heart of
the cluster Dynamical Mean-Field methods, and to realistic multiplet
structure of a correlated atom, which will allow a high precision
study of actinide and lanthanide based compounds with the combination
of the Dynamical Mean-Field theory and band structure methods.
The approach is applied to both, the two dimensional Hubbard and t-J
model within Cellular Dynamical Mean Field method. The efficient
implementation of the new algorithm, which we describe in detail,
allows us to study coherence of the system at low temperature from the
underdoped to overdoped regime. We show that the point of maximal
superconducting transition temperature coincides with the point of
maximum scattering rate although this optimal doped point appears at
different electron densities in the two models.  The power of the
method is further demonstrated on the example of the Kondo volume
collapse transition in Cerium. The valence histogram of the DMFT
solution is presented showing the importance of the multiplet
splitting of the atomic states.
\end{abstract}
\date{\today}
\maketitle

\section{Introduction}

One of the most active areas of condensed matter theory is the
development of new algorithms to simulate and predict the behavior of
materials exhibiting strong correlations.
%
%
Recent developments of the powerful many-body approach, the
Dynamical Mean Field Theory (DMFT)
\cite{old-review,phys-today,new-review} and its cluster extensions
\cite{DCA-review,CDMFT-PRL},
hold a great promise in being able to accurately predict physical
properties of this challenging class of materials.
%
In recent years, the DMFT method has substantially advanced our
understanding of the physics of the Mott transition and demonstrated
its power to explain such problems as the structural phase diagrams of
actinides \cite{Pu-nature,Am}, phonon response \cite{Pu-phonons},
optical conductivity \cite{Ce-optics}, valence and X-ray absorption
\cite{Cm-Pu} and transport \cite{transport} of some of the archetype
materials with strong correlations.

The success of the Dynamical Mean Field theory in the context of the
electronic structure revitalized the search for fast and flexible
impurity solvers which could treat Hund's coupling and spin-orbit
coupling of the parent atomic constituents in the crystal environment
of the lattice. Further, the newly developed cluster extensions of
DMFT \cite{DCA-review,CDMFT-PRL} require faster impurity solver which could
access low temperature strong correlation limit.

Many impurity solvers were developed over the last few decades,
including Hirsh-Fye Quanum Monte Carlo Method \cite{HFQMC}, exact
diagonalization \cite{ED}, non-crossing approximation \cite{NCA}, and
its extensions like one-crossing approximation
\cite{new-review,Pruschke} or SUNCA \cite{SUNCA}, iterative
perturbation theory \cite{IPT}, Wilson's numerical renormalization
group \cite{NRG} expansion around the atomic limit \cite{XDai}, and
many others.

Each method has some advantages and disadvantages, but at the present
time, there is no method that works efficiently and produces accurate
solutions for the Green’s function in all regimes of
parameters. Concentrating only on the most often employed method,
numerically exact Hirsch-Fye Quanum Monte Carlo, the following
weaknesses limit its usefulness in many realistic materials and
clusters of strongly correlated models:
%
%
%
%
i) It can not treat realistic multiplet structure which is very
important in strongly correlated $f$ materials. ii) the discretization
of the imaginary time leads to considerable systematic error
\cite{Werner_3} and requires extrapolation to infinitesimally small
time slices.  iii) the low temperatures regime in the strong
correlation limit of large $U$ is computationally very expensive since
it requires many time slices and infinite $U$ models like t-J model
are unaccessible.

All of the above mentioned shortcomings of the conventional Quantum
Monte Carlo algorithm are eliminated by the novel Continuous Time
Quantum Monte Carlo Method
\cite{Werner_1,Werner_2,Werner_3,Werner_1BM}. In addition, the new
method is even much faster for most of applications we tested,
including cluster DMFT for the Hubbard model and application of
LDA+DMFT to the Actinides.

The basic idea and its implementation for the Hubbard model was
recently published in Ref.~\onlinecite{Werner_1}. Further extensions to more
general impurity model was implemented in Ref.~\onlinecite{Werner_2}. First
demonstration of the power of our implementation was presented in
Ref.~\onlinecite{Chris} by detail low temperature study of sodium doped
cobaltates and in Ref.~\onlinecite{Cm-Pu} to study Plutonium valence.

Here we want to describe the powerful new implementation
of the method for applications to realistic materials with complicated
multiplet structure, including all interaction terms: Hubbard $U$,
Hund's coupling and spin-orbit coupling. The extension to clusters of
few sites and superconducting state within cluster DMFT will be
addressed and its power will be demonstrated by studying the low
temperature coherence of both the Hubbard and the t-J model.
The Kondo volume collapse transition in elemental Cerium will be
reexamined showing the valence histogram of the alpha and gamma phase
of the material.
More technical details of the implementation are given in the Appendix \ref{AppA}.


\section{Formalism}

In this section, we will explain the main steps of the recently developed
Quantum Monte Carlo Method with emphasis on the generalization to clusters
and multiplets of real materials.

The cluster Dynamical Mean Field approach can be conveniently 
expressed by the functional of the local Green's function \cite{new-review}
\begin{eqnarray}
  \Gamma[G_{loc}] = \Tr\log(G_0^{-1}-\Sigma)-\Tr[\Sigma G] + \Phi[G_{loc}].
\end{eqnarray}
Here $G_{loc}$ stands for the Green's function of the cluster or
single site under consideration and is in general a matrix of the size
equal to the number of sites times the number of correlated orbitals
per site. Further, $G_0$ is the non-interacting Green's function
$G_0^{-1} = \omega+\mu+\nabla^2+V_{ext}$ and contains all quadratic
terms of the Hamiltonian including periodic ionic potential of the
crystal ($V_{ext}$). The interacting part of the functional
$\Phi[G_{loc}]$ contains \textit{all} two particle irreducible
skeleton diagrams inside the chosen cluster, i.e., $\Sigma=\delta
\Phi[G_{loc}]/\delta G$.

Within the DMFT method, the summation of all the diagrams is achieved by solving
the corresponding quantum impurity problem
\begin{eqnarray}
Z=\int D[\psi^\dagger\psi] e^{-S_{c} -
   \int_0^\beta d\tau\int_0^\beta d\tau' \sum_{\a\a'}
   \psi_\a^\dagger(\tau)\Delta_{\a\a'}(\tau,\tau')\psi_{\a'}(\tau')}
 \label{Zimp}
\end{eqnarray}
where the Anderson impurity model hybridization $\Delta$ term
plays the role of the generalized Weiss field that needs to be added
to the cluster effective action $S_{c}$ to obtain the local
Green's function. The self-consistency condition which determines
this Weiss field (hybridization $\Delta$) is
\begin{eqnarray}
 G_{imp} = \frac{1}{\omega-E_{imp}-\Sigma-\Delta}=G_{loc}=\sum_\vk \frac{1}{{G_{0}^{-1}}(\vk)-\Sigma}
\end{eqnarray}

In the general impurity problem defined in Eq.~(\ref{Zimp}) the
electrons in the cluster $S_{c}$ hybridize with a matrix of
Weiss fields $\Delta_{\a\b}$.  In the case of single-site DMFT for an
$f$ material, this hybridization is a $14\times 14$ matrix while in
cluster DMFT for plaquette, it is $8\times 8$ matrix. In some cases,
hybridization can be block diagonalized as we will show in the
appendix~\ref{AppA} below on the
example of Cellular-DMFT for the normal state of the Hubbard
model. However, in the superconducting state of the same model, the
hybridization acquires off-diagonal components due to anomalous
components of self-energy.

The cluster part of the action $S_{c}$ can be very complicated
and the power of this method is that it can treat arbitrary
interaction within the cluster. For real materials study, the most
important on-site terms are the Hund's couplings which are usually
expressed by the Slater integrals \cite{Slater-paper} ($F_2$, $F_4$, $F_6$
in case of $f$ electrons). In addition, there is spin-orbit coupling
$H_{SO}=\xi\;\mathbf{l}\cdot\mathbf{s}$ and crystal field splittings
as well as various hoppings and non-local interactions within the
cluster.

The continuous time impurity solvers are based on the diagrammatic
expansion of the partition function and stochastic sampling of the
diagrams. Two expansions were recently implemented: the expansion
around the band limit \cite{Rubtsov} and the expansion in the hybridization
strength \cite{Werner_1}. The later seems to be superior in the strongly
correlated regime due to substantial reduction of the size of matrices
that need to be manipulated \cite{Werner_3} and, most importantly, the
empirical finding is that the minus sign problem in this approach is
severely reduced or maybe even eliminated \cite{Werner_2,Werner_3}.

The expansion in hybridization strength has a long history starting
with the famous Non-Crossing Approximation \cite{NCA} and
various extensions of it such as OCA \cite{new-review,Pruschke}, CTMA
\cite{CTMA}, SUNCA \cite{SUNCA}. All these approximations can be viewed as
partial summation of the same type of diagrams. With stochastic
sampling, the summation of essentially all the diagrams is now
possible. The only weakness of the approach is that it works
exclusively with the imaginary time Green's functions and analytic
continuation to real axis and access to real axis self-energy, for
example, is still hard to achieved. However, we believe that the
substantially enhanced precision of the method, as compared to
Hirsh-Fye QMC, will make this step easier.

The idea of expanding partition function in terms of the hybridization
with the conduction band, dates back to work of G. Yuval and
P.W. Anderson \cite{Anderson}. In this work, mapping the hybridization
expansion to Coulomb gas model lead to one of the first breakthroughs
in the area of the Kondo problem.  Generalization to asymmetric
Anderson model was published by H.D.M. Haldane in
Ref.~\onlinecite{Haldane}.  Similar approach was later used in
exploring the physics of the generalized Hubbard model in the context
of DMFT \cite{Si-Gabi}.  In this work, the hybridization expansion of the
partition function was analyzed by renormalization-group analysis
technique.
An early implementation of the related idea to sum up the terms of
the partition function by Monte Carlo sampling was implemented in
Ref.\onlinecite{Cox} to solve the two two-channel Anderson impurity model.

\textit{Sampled Partition Function:}
In the new Quantum Monte Carlo method, the partition function is
expanded in terms of hybridization strength $\Delta$ and the resulting
diagrams are summed up by stochastic Metropolis sampling.
Taylor expansion of the impurity partition function Eq.~(\ref{Zimp}) gives
\begin{widetext}
\begin{eqnarray}
  Z=\int D[\psi^\dagger\psi] e^{-S_{c}}\sum_k \frac{1}{k!}
  \left[
    \sum_{\a\a'}
    \int_0^\beta d\tau\int_0^\beta d\tau'
    \psi_{\a'}(\tau')\psi^\dagger_\a(\tau)\Delta_{\a\a'}(\tau,\tau')
  \right]^k
\end{eqnarray}
\end{widetext}
by separating the cluster contribution from the bath contribution the
partition function can be cast into the form
\begin{widetext}
\begin{eqnarray}
Z =\int D[\psi^\dagger\psi] e^{-S_{c}}\sum_k \frac{1}{k!}
  \int_0^\b \prod_{i=1}^k d\tau_i \int_0^\b \prod_{i=1}^k d\tau_i'
  \sum_{\a {\a}'}\prod_{i=1}^k\left[\psi_{{\a}'_i}(\tau_i')\psi_{\a_i}^\dagger
  (\tau_i)\right]
  \times \prod_{i=1}^k \Delta_{\a_i{\a}'_i}(\tau_i,\tau_i').
\label{Z_f}
\end{eqnarray}
\end{widetext}
It becomes clear that partition function is a product of two terms: the
average over the cluster states $\psi$ and average over the bath
degrees of freedom $\Delta$. It was pointed out in Ref.~\onlinecite{Werner_1} that
naive sampling of the above diagrams would run into a very bad minus
sign problem. The reason is that crossing diagrams (vertex corrections
to famous Non-Crossing Approximation to Anderson Impurity Model) can
have either sign and thus weights that correspond to crossing diagrams
could be negative. The ingenious idea proposed in Ref.~\onlinecite{Werner_1} is to
combine all the diagrams of the same order, crossing and non-crossing,
into a determinant. Mathematically, this can be expressed by
\begin{widetext}
\begin{eqnarray}
Z = Z_{c}\sum_k \frac{1}{k!}
  \int_0^\b d\tau_1 \int_0^\b d\tau_1'
  \cdots
  \int_0^\b d\tau_k \int_0^\b d\tau_k'
  \sum_{\a_1\a'_1,\cdots,\a_k,\a'_k}
 \langle T_\tau
 \psi_{\a'_1}(\tau_1')\psi_{\a_1}^\dagger (\tau_1)
 \cdots
 \psi_{\a'_k}(\tau_k')\psi_{\a_k}^\dagger (\tau_k)
 \rangle_{cluster}\times\nonumber\\
  \times \frac{1}{k!}
  Det\left(
   \begin{array}{cccc}
    \Delta_{\a_1\a'_1}(\tau_1,\tau_1') &  \Delta_{\a_1\a'_2}(\tau_1,\tau_2')&  \cdots & \cdots \\
    \cdots &\cdots & \cdots &  \cdots \\
    \cdots &\cdots & \cdots &  \cdots \\
    \Delta_{\a_k\a'_1}(\tau_k,\tau_1') & \cdots &\cdots & \Delta_{\a_k\a'_k}(\tau_{k},\tau_{k}')
   \end{array}
   \right).
   \label{Z_sample}
\end{eqnarray}
\end{widetext}
where $Z_{c}=\int D[\psi^\dagger\psi] e^{-S_{c}}$ and average of the
operator is $\langle O\rangle_{cluster}=\frac{1}{Z_{c}}\int
D[\psi^\dagger\psi] e^{-S_{c}} O$.  This is the central equation of
the Continuous Time Monte Carlo sampling around the atomic limit.  To
derive Eq.~(\ref{Z_sample}) from Eq.~(\ref{Z_f}) one needs to permute
time integration variables $\tau$ in all possible ways. Permutation of
fermions gives minus sign in an odd permutation. This minus sign can
be absorbed in the minus sign of the product of hybridizations,
resulting in the determinant of hybridizations.

\textit{Simulation:} The set of diagrams, which are associated with
the set of imaginary times
$\{\tau_1,\tau_1',\tau_2,\tau_2',\cdots,\tau_k,\tau_{k}' \}$ and
corresponding band indexes $\{\a_1,\a_1',\a_2,\a_2',\cdots,\a_k,\a_k'
\}$ are visited by Monte Carlo (Metropolis) algorithm with the weights
given by Eq.~(\ref{Z_sample}).  The effect of the hybridization
$\psi(\tau')\psi^\dagger(\tau)\Delta(\tau-\tau')$ is to create a kink
in the time evolution of the cluster, i.e., to destroy one electron at
time $\tau'$ on the cluster and create another electron at some other
time $\tau$ on the cluster. The number of kinks is always even due to
particle number conservation.

Two Monte Carlo steps which need to be implemented are: i) insertion
of two kinks at random times $\tau_{new}$ and $\tau_{new}'$ (chosen
uniformly $[0,\beta)$), corresponding to a random baths $\alpha$,
$\alpha'$.  ii) removal of a two kinks by removing one creation
operator and one annhilation operator.  Many other steps can greatly
reduce the sampling time, for example displacing a randomly chosen
operator (either $\psi$ or $\psi^\dagger$) to a new location chosen
uniformly $[0,\beta)$. The double step of inserting or removing two
kinks is also possible and is relevant when off-diagonal components of
$\Delta$ are dominant.

The detailed balance condition requires that the probability to insert
two kinks at random times $\tau$, $\tau'$, being chosen uniformly in
the interval $[0,\beta)$, is
\begin{equation}
P_{add} = min\left[\left(\frac{\beta\, N_b}{k+1}\right)^2
\frac{{\cal Z}_{new}}{{\cal Z}_{old}}\frac{{\cal D}_{new}}{{\cal D}_{old}},1\right]
\label{Padd}
\end{equation}
where $N_b$ is the number of baths, $k$ is the current
perturbation order (number of kinks/2), ${\cal Z}_{new}$ is the
cluster matrix element
\begin{eqnarray}
{\cal Z}_{new}=\langle T_\tau
\psi_{\a'_{new}}(\tau'_{new})\psi^\dagger_{\a_{new}}(\tau_{new})
\psi_{\a'_1}(\tau_1')\psi_{\a_1}^\dagger (\tau_1) \cdots\nonumber\\
\cdots\psi_{\a'_k}(\tau_k')\psi_{\a_k}^\dagger (\tau_k) \rangle_{cluster}
\end{eqnarray} 
 and ${{\cal D}_{new}}/{{\cal D}_{old}}$ is the ratio between the
 new and the old determinant of baths $\Delta$ and can be evaluated using
 usual Shermann-Morrison formulas. The factors of $(\beta N_b)$ enter
 because of the increase of the phase space when adding a kinks
 (increase of entropy) while the factor $1/(k+1)$ comes from
 factorials in Eq.~\ref{Z_sample}.
Similarly, the probability to remove two kinks, chosen randomly
between $[1\cdots k]$ is
\begin{equation}
P_{remove} = min\left[\left(\frac{k}{\beta\, N_b}\right)^2
\frac{{\cal Z}_{new}}{{\cal Z}_{old}}\frac{{\cal D}_{new}}{{\cal D}_{old}},1\right].
\label{Premove}
\end{equation}
 

An important simplification occurs if the hybridization is block
diagonal. Since $\Delta_{\a\a'}$ vanishes for some combination of
$\a\a'$ the determinant in Eq.~(\ref{Z_sample}) can be written as a
product of smaller determinants, one for each block of hybridization.
A specially simple case occurs when all the blocks are of size one,
the partition function becomes a product of $N_b$ terms, where $N_b$ is
number of all baths
\begin{widetext}
\begin{eqnarray}
Z = Z_{c}
\sum_{\{k_\a\}} \frac{1}{k_\a!}
  \int_0^\b \prod_{\a=1}^{N_b}\prod_{i=1}^{k_\a} d\tau_i^\a \int_0^\b
  \prod_{\a=1}^{N_b}\prod_{i=1}^{k_\a} d{\tau'}_i^{\a}
 \langle T_\tau
 \prod_{\a}
 \psi_{\a}({\tau'}_1^\a)\psi_{\a}^\dagger (\tau_1^\a)
 \cdots
 \psi_{\a}({\tau'}_{k_\a}^\a)\psi_{\a}^\dagger (\tau_{k_\a}^\a)
 \rangle_{cluster}\times\nonumber\\
  \times \prod_\a \frac{1}{k_\a!}
  Det\left(
   \begin{array}{cccc}
    \Delta_{\a}(\tau_1^\a,{\tau'}_1^\a) &  \Delta_{\a}(\tau_1^\a,{\tau'}_2^\a)&  \cdots \\
    \cdots &\cdots & \cdots \\
    \cdots &\cdots & \cdots \\
    \Delta_{\a}(\tau_{k_\a},{\tau'}_{k_\a}) & \cdots &\Delta_{\a}(\tau_{k_\a}^\a,{\tau'}_{k_\a}^\a)
   \end{array}
   \right)
\label{Z1b1}
\end{eqnarray}
\end{widetext}
The probability to add two new kinks Eq.~(\ref{Padd}) now depends on the
number of kinks of this particular type $k_\a$ (rather than total
perturbation order $k$) and the dimension of the bath subspace $N_b$
becomes unity. In general the
probability to add two kinks is
$P_{add} = \left(\frac{\beta\, N_b^\a}{k_\a+1}\right)^2
\frac{{\cal Z}_{new}}{{\cal Z}_{old}}\frac{{\cal D^\a}_{new}}{{\cal D^\a}_{old}}$,
where $N_b^\a$ is the number of bands which form
an off-diagonal sub-block in hybridization $\Delta$ and need to be
treated simultaneously in one determinant ${\cal D^\a}$ in Eq.~(\ref{Z_sample}).

The size of hybridization determinants can thus be greatly reduced if
hybridization $\Delta$ is block diagonal.  The cluster term ${\cal Z}$
however cannot be broken into separate contributions for each bath,
rather the full trace needs to be computed numerically. It is
therefore essential to find a fast way to compute the cluster average
$\langle\cdots\rangle_{cluster}$ in
Eq.~(\ref{Z_sample}).

\textit{Exact Diagonalization of the Cluster:} It is crucial to
evaluate the cluster trace in eigenbase of the cluster Hamiltonian and
take into account the conservation of various quantum numbers, such as
the particle number, the total spin, and the total momentum of the
cluster states.

Typical contribution to the cluster part of the trace that needs to
be evaluated at each Monte Carlo step, takes the form
\begin{widetext}
\begin{eqnarray}
\label{evolution}
&&
{\cal Z_D}=
\Tr\left(T_\tau e^{-\int_0^\b d\tau H_{c}(\tau)}
\psi_{\a_1}(\tau_1')
\psi^\dagger_{\a_2}(\tau_2)
\cdots
\psi_{\a_{n-1}}(\tau_{n-1}')
\psi^\dagger_{\a_n}(\tau_n)
\right)\\
&&=
\sum_{\{m\}}
e^{-E_{m_1}\tau_1'}
(F^{\a_1})_{m_1 m_2}
e^{-E_{m_2}(\tau_2-\tau_1')}
(F^{\dagger\a_2})_{m_2 m_3}
\cdots
(F^{\a_{n-1}})_{m_{n-1} m_{n}}
e^{-E_{m_n}(\tau_{n-1}'-\tau_{n})}
(F^{\dagger\a_n})_{m_n m_1}
e^{-E_{m_1}(\beta-\tau_n)}
\nonumber
\end{eqnarray}
\end{widetext}
where the matrix elements are $(F^{\dagger \a_i})_{n m}= \langle
n|\psi^\dagger_{\a_i}|m\rangle$ and $E_m$ are eigenvalues of the
cluster. The actual order of operators in Eq.~(\ref{evolution})
depends on their time arguments and creation operator is not necessary
followed by annhilation operator.

The bottleneck of the approach is that the number of cluster states
$|m\rangle$ is very large (for example, single site DMFT for the $f$
shell requires 2$^{14}$ states). Consequently, the matrix elements $F$
are in general very large matrices and the typical diagram order is
inversely proportional to temperature (see Fig.~\ref{Fig1}) therefore
one typically needs to multiply few hundred large matrices at each Monte Carlo
step. It is clear that this is very impractical and the progress can
be achieved only by various trick which we have implemented
\begin{itemize}
\item Most of matrix elements vanish. A fast algorithm is needed to
  determine which matrix elements are nonzero.
\item Symmetries of the problem can  be taken into account to reduce
  the size of the $F$ matrix.
\item The number of trial steps is usually much bigger (100 times)
  than the number of accepted steps and the insertion or removal of a
  kink is very local in time operation.  It is convenient to store the
  product Eq.~\ref{evolution} (from both sides, left and right) and
  when trying to insert a new kink, recompute the trace only between
  the inserted times $\tau_{new}$ and $\tau_{new}'$.
  
\item During simulation, the probability for visiting any cluster
  state can be recorded and can be used in the next step to eliminate
  the irrelevant atomic states from the trace in
  Eq.~\ref{evolution}. The cluster base can hence be adjusted
  dynamically to describe the particular regime studied by the minimum
  number of relevant cluster states.
\end{itemize}

To illustrate the method, let us consider a concrete example of the
cluster of one band model (Hubbard or t-J) in the normal state. The
bath index $\alpha$ runs over cluster momenta $\vq$ and spin
orientation $\sigma$. The eigenstates of the cluster can be written in a
form $|N,S_z,\vK;S\gamma\rangle$, where $N$ is total number of
electrons in the state, $S$ and $S_z$ are the total spin and its $z$
components, while $\vK$ is the total momentum of the cluster state and
$\gamma$ stands for the rest of the quantum numbers.

\textit{Concept of Superstates:}
In this base, the matrix elements of the creation operator are greatly
simplified $\psi^\dagger_{\vq,\sigma}|N,S_z,\vK;S\gamma\rangle =
|N+1,S_z+\sigma,\vK+\vq;S\pm 1/2,\gamma\rangle$ because the creation
operator is nonzero only between Hilbert subspace of $\{N,S_z,\vK\}$
and $\{N+1,S_z+\sigma,\vK+\vq\}$. It is therefore convenient to group
together states with the same $\{N,S_z,\vK\}$ and treat the rest of
the quantum numbers as internal degrees of freedom of a cluster
superstate $|i\rangle\equiv|\{N,S_z,\vK\}\rangle$. The superstate
$|i\rangle$ is multidimensional state with internal quantum numbers
$|m[i]\rangle\equiv|\{S,\gamma\rangle\}$.  It is then clear that creation
operator acting on a state $|i\rangle$ gives a unique state $|j\rangle
= \psi^\dagger_{\vq\sigma}|i\rangle$ and it is enough to store a
single index array $F^{\a\dagger}(i)=j$ to figure out how the Hilbert
subspaces are visited under application of a sequence of creation and
annhilation operators such as in Eq.~(\ref{evolution}):
$i_0\rightarrow F^{\a_1}(i_0)\rightarrow \cdots
i_k=F^{\dagger\a_k}(i_{k-1})$. This sequence is very often truncated
in few steps only, because most of the sequences contain either multiple
application of the same creation or annhilation operator (Pauli
principle) or because they lead to a state outside the base
(for example $\psi |N=0...\rangle=0$ or
$\psi^\dagger|N=max...\rangle=0$).

Once the nonzero matrix elements are found by this simple index
lookup, the value of the matrix element needs to be computed by matrix
multiplication. By this breakup of the Hilbert space and introducing
superstates $|i\rangle$, the largest matrix which needs to be treated
for the t-J model on a plaquette is $3\times 3$ and for the Hubbard
model it is $6\times 12$, thus substantially smaller than the original
$81$ and $265$ dimensional Hilbert space. To compute the trace in
Eq.~(\ref{evolution}) we start with unity matrix in each subspace of a
superstate $|i\rangle$ and apply both the time evolution operator
$e^{-E_m(\tau_l-\tau_l')}$ (by multiplying each row of a matrix with
its time evolution) and the kink (by multiplication with the matrix
$(F^{\alpha})_{mn}$ or $(F^{\alpha\dagger})_{mn}$). The operation of
$F$ brings us to the next superstate $F^\a(i)$ where we repeat both
the time evolution and the kink application. After $k$ steps, the
trace of a matrix gives the desired matrix element.

\textit{Storing the Time Evolution:} The number of kinks is
proportional to inverse temperature $\beta$ and kinetic energy of the
system $\langle k\rangle= \beta |E_{kin}|$ (see Eq.~(\ref{avek})). It
thus becomes large at low temperatures. However, an insertion of a
kink with large time difference ($\psi^\dagger(\tau)\psi(\tau')$ with
large $\tau-\tau'$) has a very low probability.  The reason is that
Pauli principle forbids insertion of the pair
$\psi_\a^\dagger(\tau)\psi_\a(\tau')$ if another kink of the same
species $\a$ is between the two times $(\tau,\tau')$. At the same
time, $\Delta(\tau)$ is like $G(\tau)$ peaked at small times
$\tau-\tau'$ and falls of at large times making the long time
intervals rare.

The insertion of a kink is thus fairly local in time
operation, therefore it is convenient to store a whole chain of
products that appear in Eq.~(\ref{evolution}) from both sides, left
and right, to make trial step very cheap. It takes only few matrix
multiplications (almost independent of temperature) to compute the
trace in Eq.~(\ref{evolution}). When the move is accepted, the trace
needs to be updated which takes somewhat more time. However, the
acceptance rate is typically small and on average does not require
much computational time.

\textit{Adjusting the Cluster Base:} Finally, the ultimate speedup can
be achieved by dynamically adjusting "the best" cluster base. The
probability for a cluster state $|m\rangle$ can be computed during
simulation. For a given diagram with particular configuration of
kinks, the probability for a cluster state $|m\rangle$ is proportional
to its matrix element defined in Eq.~(\ref{evolution}), i.e.,
\begin{eqnarray}
  P_m = \frac{\langle m|e^{-H\tau_1} \psi_{\a}
    e^{-H(\tau_2-\tau_1)}\cdots|m\rangle}
  {\sum_{\{n\} \langle n|e^{-H\tau_1} \psi_{\a} e^{-H(\tau_2-\tau_1)}\cdots|n\rangle}}.
\label{Pm}
\end{eqnarray}
The sampled average of this quantity gives probability for cluster
eigenstate $|m\rangle$. A large number of cluster states have very
small probability and can be eliminated in simulation to ultimately
speed up the simulation. It is important that the probability for 
any cluster state depends on the particular problem at hand and the
program adjusts the base dynamically after a few million of Monte Carlo
steps.

\textit{Green's function Evaluation:} Like in other impurity solvers which are based on the expansion of the
hybridization (NCA, OCA, SUNCA), the Green's function is computed from
the bath electron T matrix. Using equation of motion, it is easy to
see that the bath Green's function $G_{kk'}$ is connected to the local
Green's function $G_{loc}$ through the following identity
\begin{eqnarray}
&&G_{kk'}(\tau-\tau') = \delta_{kk'} g_{kk}(\tau-\tau')\\
&&+ \int_0^\beta \int_0^\beta d\tau_s d\tau_e\, g_{k}(\tau-\tau_e)V_k G_{loc}(\tau_e-\tau_s)V_{k'}g_{k'}(\tau_s-\tau')\nonumber
\end{eqnarray}
or summing over momenta
\begin{eqnarray}
  {\cal G}(\tau-\tau') \equiv \sum_{kk'}V_k G_{kk'}(\tau-\tau')V_{k'} = \Delta(\tau-\tau')\nonumber\\
  + \int_0^\beta \int_0^\beta d\tau_s d\tau_e\, \Delta(\tau-\tau_e)
  G_{loc}(\tau_e-\tau_s) \Delta(\tau_s-\tau').
\label{Gl1}
\end{eqnarray}
This Green's function ${\cal G}$ is equal to the ratio between the
determinant of $\underline{\Delta}$'s
\begin{eqnarray}
 {\cal G}(\tau-\tau') = \frac{
Det\left(
   \begin{array}{l|rrr}
   \Delta(\tau,\tau') & \Delta(\tau,{\tau'}_1) & \cdots & \\
    \hline
    \Delta(\tau_1,\tau') & &  & \\
    \cdots & \underline{\Delta} & & \\
   \Delta(\tau_{k},\tau')& & &
   \end{array}
   \right)
   }{
Det\left(\underline{\Delta}\right)
   }\nonumber
\end{eqnarray}
where one row and one column is added to the bath electron
determinant. The reason for this simple form is that the conduction
band is non-interacting and thus obeys Wicks theorem.
Here we used the definition
\begin{equation}
 \underline{\Delta}\equiv M^{-1} = \left(
   \begin{array}{cc}
    \Delta_{\a}(\tau_1^\a,{\tau'}_1^\a)&  \cdots \\
    \cdots & \cdots \\
    \cdots &\Delta_{\a}(\tau_{k_\a}^\a,{\tau'}_{k_\a}^\a)
   \end{array}
   \right)
\end{equation}
Using Shermann Morrison formulas to express enlarged determinant by
the original determinant of $\underline{\Delta}$, ${\cal G}$ becomes
\begin{eqnarray}
  {\cal G}(\tau-\tau') = \Delta(\tau-\tau') - \sum_{i_e,i_s}\Delta(\tau-\tau_{i_e})M_{i_e,i_s}\Delta(\tau_{i_s}-\tau')
\label{Gl2}
\end{eqnarray}
Finally, comparing Eq.~(\ref{Gl1}) and (\ref{Gl2}) we see that
$ G_{loc}(\tau_e-\tau_s) = -M_{i_e,i_s}$ and in imaginary frequency
\begin{eqnarray}
  G_{loc}(i\omega) = -\frac{1}{\beta}\sum_{i_e,i_s}e^{i\omega\tau_{i_e}}M_{i_e,i_s} e^{-i\omega\tau_{i_s}}.
\end{eqnarray}
This equation is the central equation of the approach since it relates
local green's function with the quantities computed in QMC importance
sampling. The equation shows that only matrix
$M \equiv (\underline{\Delta})^{-1}$ needs to be stored and
manipulated in simulation.

From the above consideration, it is clear that one could also add two
rows and two columns to matrix $\underline{\Delta}$ and compute the
two particle vertex function in similar way without much overhead.

\textit{Fast Updates:} The Green's function can be updated in linear time (rather than
quadratic). When adding a construction and annhilation operator at
$\tau$ and $\tau'$, adding a column at $i_s$ and row at $i_e$ to
matrix $M^{-1}$, leads to the following relation between the matrix
$M_{new}$ and $M_{old}$
\begin{equation}
  M^{new}_{ji}=M^{old}_{ji}+ p L_{j}R_{i}.
\end{equation}
Here one row and one column of zeros is added to $M^{old}$ to match
the size of $M^{new}$. The arrays $L$ and $R$ are given by
\begin{eqnarray}
 L = (\widetilde{L}_1,\cdots,\widetilde{L}_{i_e-1},-1,\cdots,\widetilde{L}_k)\\
 R = (\widetilde{R}_1,\cdots,\widetilde{R}_{i_s-1},-1,\cdots,\widetilde{R}_k)
\end{eqnarray}
where
\begin{eqnarray}
  \widetilde{L}_j = \sum_i M_{ji}^{old}\Delta(\tau_i-\tau')\\
  \widetilde{R}_i = \sum_j \Delta(\tau-\tau_j)M_{ji}^{old}
\end{eqnarray}
and $p$ is
\begin{equation}
 \frac{1}{p} = \Delta(\tau-\tau')-\sum_{ij}\Delta(\tau-\tau_j)M_{ji}^{old}\Delta(\tau_i-\tau').
\end{equation}
Finally, the local Green's function becomes
\begin{equation}
  G^{new} = G^{old} - \frac{p}{\beta}
  \left(\sum_{j_e} e^{i\omega\tau_{j_e}} L_{j_e}\right)
  \left(\sum_{j_s}R_{j_s} e^{-i\omega\tau_{j_s}}\right).
  \label{dGn}
\end{equation}
It is clear from Eq.~(\ref{dGn}) that only linear amount of time is
needed to update the local Green's function.

When removing two kinks of construction and annhilation operators at
$i_s$ and $i_e$, old and new matrix $M$ are related by
\begin{equation}
  M^{new}_{ij}=M_{ij}^{old} - \frac{M_{i i_s}^{old}M_{i_e j}^{old}}{M_{i_e,i_s}^{old}}
\end{equation}
Green's function therefore becomes
\begin{equation}
  G^{new} = G^{old} + \frac{1}{\beta M_{i_e i_s}}
  \left(\sum_{j_e} e^{i\omega\tau_{j_e}}  M_{j_e  i_s}\right)
  \left(\sum_{j_s}M_{i_e j_s} e^{-i\omega\tau_{j_s}}\right).
\end{equation}

Finally, the exponential factors $e^{i\omega\tau_i}$ do not need to be
recomputed at each Monte Carlo step since all "old" times can be
stored and the exponents need to be computed only for the new pair of
times and only at each accepted move. In the present implementation,
the algorithm to sample directly $G(i\omega)$ is sufficiently fast
that it does not introduce any performance costs. Since it does not
introduce systematic error in binning $G(\tau)$ we believe it is
superior than the alternative implementations which sample $G(\tau)$.

\textit{Large Frequencies and Moments:}
Similarly to the Hirsch-Fye QMC, the low frequency points of Green's
function converge very fast to the exact value while the
high-frequency points, when sampled directly, contain a lot of
noise. It is therefore not very useful to sample large frequencies in
the above described way.  Usually we sample 200-300 Matsubara points
while the rest are replaced by the high frequency moments of
the self-energy computed analytically.

The high frequency moments of the self-energy are computed from the
Green's function moments, which, in general, take the following form
\begin{equation}
m_n^{\a\b} = (-1)^n\left\langle \left\{ [H,[H,\cdots[H,\psi_\a]\cdots]], \psi^\dagger_\b\right\}\right\rangle
\end{equation}
To compute the moments within the present approach, few operators need
to be sampled in simulation. In the one-band model, only density is
required, but in more complicated situation, higher order
density-density and exchange terms enter. In general, Green's function
moments can be expressed by the average of a few equal time operator
$O_{mm'}$, which take particular simple form in the local eigenbase
\begin{eqnarray}
\label{Gmoments1}
m^{G,\a\b}_{1,mm'} = \sum_n (F^\a)_{mn} (F^{\b\dagger})_{nm'}(E_n-E_m)\\
                   - (F^{\a\dagger})_{mn}(F^{\b})_{nm'}(E_{n}-E_{m'})\nonumber\\		   
\label{Gmoments2}
m^{G,\a\b}_{2,mm'} = \sum_n (F^\a)_{mn} (F^{\b\dagger})_{nm'}(E_m-E_n)^2\\
                   +(F^{\a\dagger})_{mn}(F^{\b})_{nm'}(E_{n}-E_{m'})^2\nonumber .
\end{eqnarray}
The lowest order self-energy moments can then be computed in the
following way
\begin{eqnarray}
&&  \Sigma_{\a\b}(\infty) = \langle m^{G,\a\b}_{1}\rangle - E_{imp}\\
&&  \Sigma^{(1)}_{\a\b} = \langle m^{G,\a\b}_{2}\rangle - \langle m^{G,\a\b}_1\rangle^2.
\end{eqnarray}

\textit{Sampling of other Quantities:}
Equal time operator like density or those in Eq.~(\ref{Gmoments1}) and
(\ref{Gmoments2}) can be straightforwardly computed in simulation. To
improve the precision of simulation, operators are averaged over all
times, i.e.,
\begin{equation}
\langle O\rangle = \frac{1}{Z}\frac{1}{\beta}\int_0^{\beta}d\tau_0
\Tr[T_\tau e^{-\int_0^{\tau_0} H d\tau}
  O(\tau_0) e^{-\int_{\tau_0}^{\beta} H d\tau}]
\end{equation}
The contribution to $\langle O\rangle$ of each particular diagram
is
\begin{equation}
\frac{1}{{\cal Z_D}}\sum_{l=1}^{k}\int_{\tau_l}^{\tau_{l+1}}d\tau_0 \,
\Tr[T^{left}_l e^{-E(\tau_0-\tau_l)} O(\tau_0) e^{-E(\tau_{l+1}-\tau_0)}T^{right}_{k-l}]
\label{O1}
\end{equation}
where ${\cal Z_D}$ is the cluster part of the trace
Eq.~(\ref{evolution}), $T^{left}_l$ are the operators in
Eq.~(\ref{evolution}) which appear before $\tau_0$ and
$T^{right}_{k-l}$ contains operator with time arguments greater than
$\tau_0$.  $k$ is the number of kinks in the diagram $\cal D$.  As we
mentioned before, these time evolution operators ($T^{left}_l$,
$T^{right}_{l}$) are stored and regularly updated during simulation.

Further simplification is possible, if operator $\langle O\rangle$
commutes with $H_{cluster}$
\begin{equation}
\langle O\rangle_{\cal D}=\frac{1}{{\cal Z_D}}
\sum_l\frac{\tau_{l+1}-\tau_l}{\beta}\,\Tr[T^{left}_l e^{-E(\tau_{l+1}-\tau_l)} O T^{right}_{k-l}].
\end{equation}
If operator $O$ is one of the good quantum numbers of a superstate
(one of $\{N,S_z,\vK\}$ in our example above), matrix $O_{m_1,m_2}$ is
proportional to identity matrix in each superstate
$O_{m_1[i],m_2[i]}=\delta_{m_1,m_2} O_i$. Starting with a superstate
$i$, kinks of a diagram $\cal D$ generate a
sequence of superstates
$i\rightarrow j_{1}=F^{\a_1}(i) \rightarrow
j_{2}=F^{\dagger\a_2}(j_{1})\cdots\rightarrow i$, as explained above.
The equal time average of cluster conserved quantity therefore simplifies to
\begin{eqnarray}
\langle O\rangle_{\cal D}=  \sum_{i\in superstates} P_i \sum_{l=1}^k \frac{\tau_{l+1}-\tau_l}{\beta} O_{j_l}
\end{eqnarray}
where $P_i$ is probability for a superstate $|i\rangle$ defined in
Eq.~(\ref{Pm}) and $O_{j_l}$ is the value of the conserved quantity in
each superstate $j_l$ in the interval $\tau_{l+1}-\tau_l$. This time
difference is the time each superstate "lives" in the particular
diagram.

In this way, the electron density $n_f$ and average magnetization
$S_z$ is computed to very high precision.  Similar simplification in
evaluating the total spin susceptibility, defined by
\begin{eqnarray}
\chi(i\omega)=\frac{1}{\beta}\int_0^\beta d\tau \int_0^\beta d\tau'
\langle S_z(\tau) S_z(\tau')\rangle
\end{eqnarray}
where $S_z$ is the total spin of the cluster (or atom) leads to
\begin{eqnarray}
\chi(i\omega)=\sum_{i\in superstates} P_i
\left|\sum_{l=1}^k
(S_z)_{j_l} \frac{e^{i\omega \tau_{l+1}}-e^{i\omega\tau_l}}{i\omega}\right|^2
\end{eqnarray}
where $(S_z)_j$ is the total spin of the cluster in cluster eigenstate
$j$.

If $O$ does not commute with $H_{c}$, the integral (\ref{O1})
needs to be evaluated.  In cluster eigenbase it takes the form
\begin{eqnarray}
&& \frac{1}{{\cal Z_D}}
\sum_{l m m_1 m_2}(T^{left}_l)_{m m_1}
\frac{e^{-E_{m_2}(\tau_{l+1}-\tau_{l})}-e^{-E_{m_1}(\tau_{l+1}-\tau_l)}}{\beta(E_{m_1}-E_{m_2})}\nonumber\\
&& \qquad\qquad\qquad\qquad\times O_{m_1 m_2} (T^{right}_{k-l})_{m_2 m}
\end{eqnarray}

\textit{Sampling the total energy:}
The average of the potential energy $\langle V\rangle$ can be computed from the average
of the local energy since
\begin{eqnarray}
\langle H_{loc}\rangle &=& \sum_{\a\b\g\d} U_{\a\b\g\d}\langle\psi_\a^\dagger\psi_\b^\dagger\psi_\d\psi_\g\rangle
+\sum_{\a\b}E_{imp\,\a\b}\langle\psi^\dagger_\a\psi_\b\rangle \nonumber\\
&=&\langle V\rangle + \Tr[E_{imp} n]
\end{eqnarray}
Here we concentrate on the case of general single site DMFT rather
than the cluster extensions. The reason is that kinetic energy in
cluster extensions depends on the periodization scheme and we will not
go into this details here.
Kinetic energy of the general single site DMFT $E_{kin}=\Tr[H^0_\vk G_\vk]$ can be computed
by
\begin{eqnarray}
  E_{kin} = \Tr[\Delta G] + \Tr[(\mu+E_{imp})n]
\end{eqnarray}
The total energy is therefore given by
\begin{eqnarray}
  \langle H\rangle= \langle H_{loc}\rangle + \Tr[\Delta G] 
  + \mu n
\end{eqnarray}
The first term $H_{loc}$ can be computed very precisely in
simulation. 
The sampled quantity $\langle O\rangle$ is just the energy of an atomic
state and can be simply obtained from the probabilities for atomic states
$\langle H_{loc}\rangle = \sum_{m\in all-states} P_m E_m$.
Computing kinetic energy from the Green's function gives in general worse
accuracy because the high-frequency behavior of the Green's function
can not be directly sampled and augmentation with analytically
computed tails is necessary.
However, it is simple to show that the average value of the
perturbation order is related to the average of the kinetic energy
\begin{equation}
 \langle k\rangle = -\frac{1}{T} \Tr[\Delta G]
\label{avek}
\end{equation}
where $\langle k\rangle$ is the average perturbation order and $T$ is
temperature. The later quantity is directly sampled in the present
algorithm and it is just the center of gravity of the histogram
(presented in Fig.~\ref{Fig1}).
Finally, the total energy $E$ is given by
\begin{eqnarray}
E_{tot} = \langle H_{local} \rangle - T \langle k\rangle + \mu n
\end{eqnarray}
All quantities in this equation can be computed to very high accuracy
and since low temperatures can be reachable in this method, the entropy can
be obtained by integrating the specific heat.

\textit{Superconductivity:} The power of the method can be further
demonstrated by studying the superconducting state of the strongly
correlated systems at low temperature with essentially no performance
cost. By employing the Nambu formalism, the translationally invariant
cluster methods (for details on Cellular-DMFT on a plaquette see
Appendix \ref{AppA}) results in $N_c$ two dimensional baths, where $N_c$ is the
number of cluster momenta. Namely, the baths $\{\vK,\uparrow\}$ and
$\{-\vK,\downarrow\}$ are coupled through the anomalous component of
hybridization and require simultaneous treatment in
$Det(\underline{\Delta})$ in Eq.~(\ref{Z1b1}). The determinants are on
average twice as large as in normal state, however,
%
the cluster part of the trace in Eq.~(\ref{Z1b1}) remains
unchanged. Since most of the time is usually spend in evaluating the
local part of $Z$, the performance is not noticeably
degraded in superconducting state. In typical run presented below, the
histogram is peaked around $k=250-500$ which is equal to the order of
a typical diagram. In the translational invariant representation
employed here, the size of a typical determinant in Eq.~(\ref{Z1b1})
is only $k/N_c\sim 60-120$ and using fast-update scheme presented
above, the trace over the bath states is not expensive part of the
algorithm.

\textit{Hund's coupling and spin-orbit coupling:}
In materials with open $f$ orbitals, the multiplet effects are very
strong and $SU(N)$ approximation is not adequate. Simultaneous
inclusion of Hund's coupling and spin-orbit coupling in DMFT method is
crucial for quantitative description of Actinides \cite{Cm-Pu}.
Minimal local Hamiltonian for Lanthanide and Actinide materials is
\begin{equation}
H_{atom} = H_{Hubbard+Hunds}+H_{SO}+\widetilde{E}_{imp} \hat{n}
\end{equation}
Here $\widetilde{E}_{imp}$ is the impurity level without the
spin-orbit coupling since later is included explicitely. The
Hund's coupling and spin-orbit coupling take the following form
\begin{widetext}
\begin{eqnarray}
&& H_{Hubbard+Hunds}=
\sum_{L_i,m,\sigma\sigma'}\sum_{k=0}^{2 l}\frac{4\pi F^{k}_{\{l\}} }{2 k+1}
\langle Y_{L_a}|Y_{k m}|Y_{L_c}\rangle
\langle Y_{L_d}|Y_{k m}|Y_{L_b}\rangle
f^\dagger_{L_a\sigma} f^\dagger_{L_b\sigma'}f_{L_d\sigma'}f_{L_c\sigma}\\
&& H_{SO}=\sum_{jm_j,lmm',\sigma\sigma'}\xi\frac{1}{2}[j(j+1)-l(l+1)-\frac{3}{4}]
C^{j m_j}_{lm,\sigma}C^{j m_j}_{l m',\sigma'}
f_{lm\sigma}^\dagger f_{lm'\sigma'}
\end{eqnarray}
\end{widetext}
Here $Y_L$ are spheric harmonics, $C^{j m_j}_{lm,\sigma}$ are
Clebsch-Gordan coefficients and $F^{k}$ are Slater integrals.
The $F^{0}$,  usual Hubbard $U$, is commonly computed by constrained
LDA, while the rest of the Slater integrals ($F^2$, $F^4$ and $F^6$) are
computed using atomic physics program \cite{Cowan}. Finally, the spin-orbit
strength $\xi$ is computed within LDA program and needs to be updated
during self-consistent LDA+DMFT calculation.

Exact diagonalization of the atomic Hamiltonian leads to
eigenstates with conserved number of particles $N$, total angular momentum
$J$ and its $z$ component $|N J_z;J\gamma\rangle$.

For the one electron base, it is more convenient to chose the electron
total angular momentum $(j,j_z)$ rather than $(l_z,\sigma)$ since the
construction and annhilation operators conserve $z$ component of total
spin
\begin{eqnarray}
\psi^\dagger_{j j_z}|N J_z; J \gamma \rangle =|N+1,J_z+j_z; J' \gamma'\rangle.
\end{eqnarray}
%
%
%
%
Superstates $|i\rangle$ are chosen in a way to simplify the
matrix elements of the creation operator $\psi^\dagger$. In this case,
a convenient choice is
\begin{equation}
  |i\rangle\equiv |\{N J_z\}\rangle.
\end{equation}
In the absence of crystal field splitting, the hybridization $\Delta$
is a diagonal matrix (all $14$ baths are one-dimensional). However,
strong crystal fields splitting generates off-diagonal components of
$\Delta_{\{jj_z\}\{j'j_z'\}}$ and in general, the determinant
$Det(\underline{\Delta})$ in Eq.~(\ref{Z_sample}) can not be broken into
small determinants as in Eq.~(\ref{Z1b1}).

\section{Results}

We implemented Cellular DMFT for both the Hubbard and the t-J model on
a plaquette. The energy scales are given in units of $t$, Hubbard $U$
is fixed at $12U$ while $J$ of the t-J model is set to $J=0.3$.  We
simulated $5.000.000$ Monte Carlo steps per processor and results were
averaged over 64 processors. One DMFT step for the Hubbard model takes
approximately 45 minutes on 1.7$\,$GHz PC processor therefore each
Monte Carlo step requires on average one million clock-ticks (0.9
Mflops).

\begin{figure}[htbp]
\includegraphics[width=0.9\linewidth]{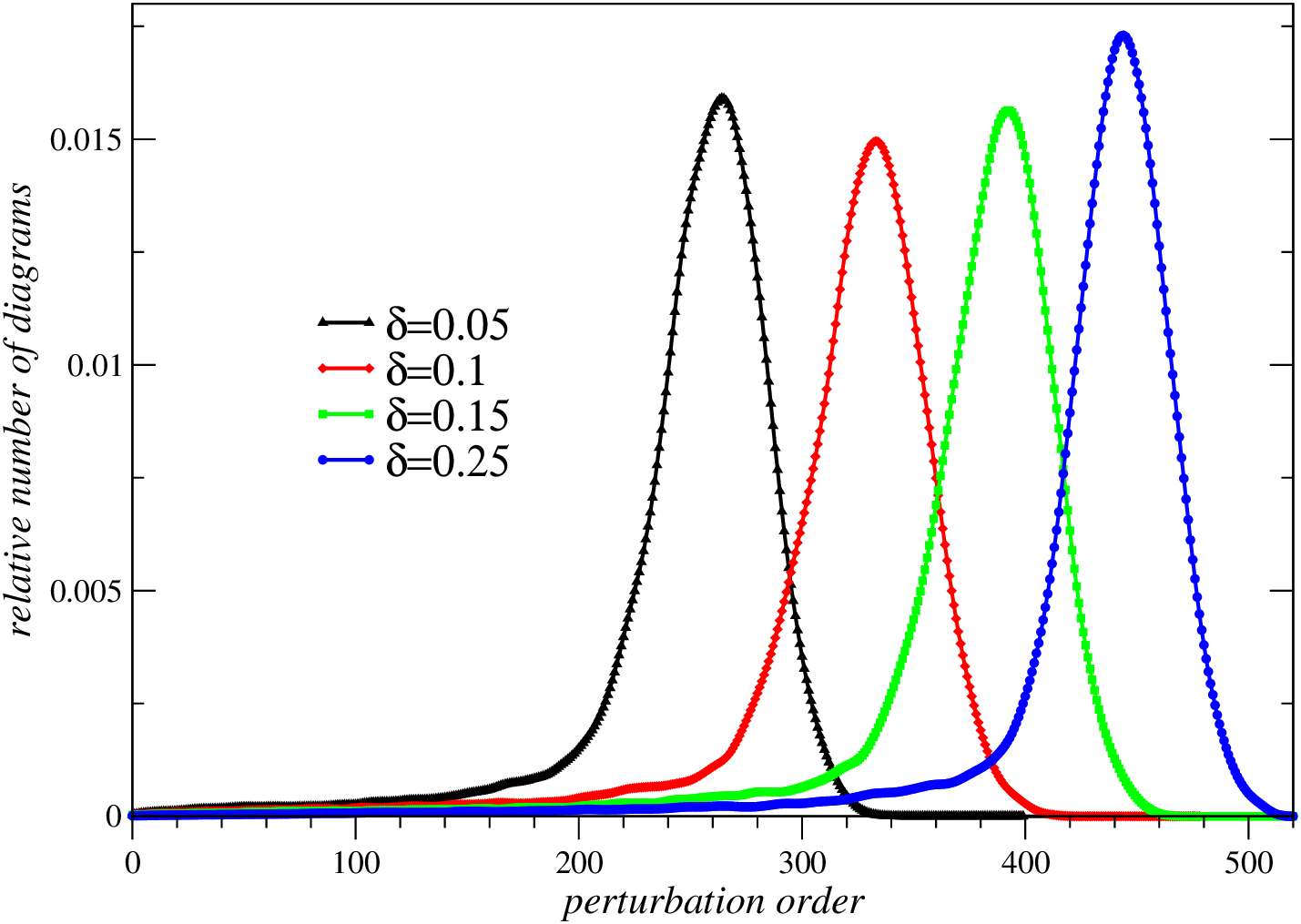}
\caption{
The \textit{perturbation order histogram} shows the
distribution of the typical perturbation order of the diagrams in the simulation.
The histogram is peaked around the typical order, which is related to
temperature and kinetic energy by $\langle k\rangle = |E_{kin}|/T$.
}
\label{Fig1}
\end{figure}
Fig.~\ref{Fig1} shows histograms (probability distribution for the
perturbation order $k$) for few dopings of the Hubbard model
$\delta=1-n$ at $T=1/200t$.  The average perturbation order is increasing
with doping since the electrons are getting more delocalized (absolute
value of the kinetic energy is increasing) and the
creation of kinks becomes less expensive. It is of the order of
450 in overdoped regime, but the typical size of the
determinant is only 112 in superconducting state or 56 in normal state.

We recently addressed the problem of coherence scale in the t-J model
\cite{QCP-paper} and we found, using NCA as the impurity solver, that
the imaginary part of the cluster self-energy $\Sigma_{(\pi,0)}$,
which plays the crucial role in the optical conductivity and
transport, becomes very large at optimal doping and consequently the
coherence scale vanishes around optimal doping. Here we extend this
study to the Hubbard model using much lower temperature.  We will show
that the system become strongly incoherent at optimal doping and the
maximum of $T_c$ tracks the maximum scattering rate in both the
Hubbard and the t-J model.

\begin{figure}[htbp]
\includegraphics[width=0.9\linewidth]{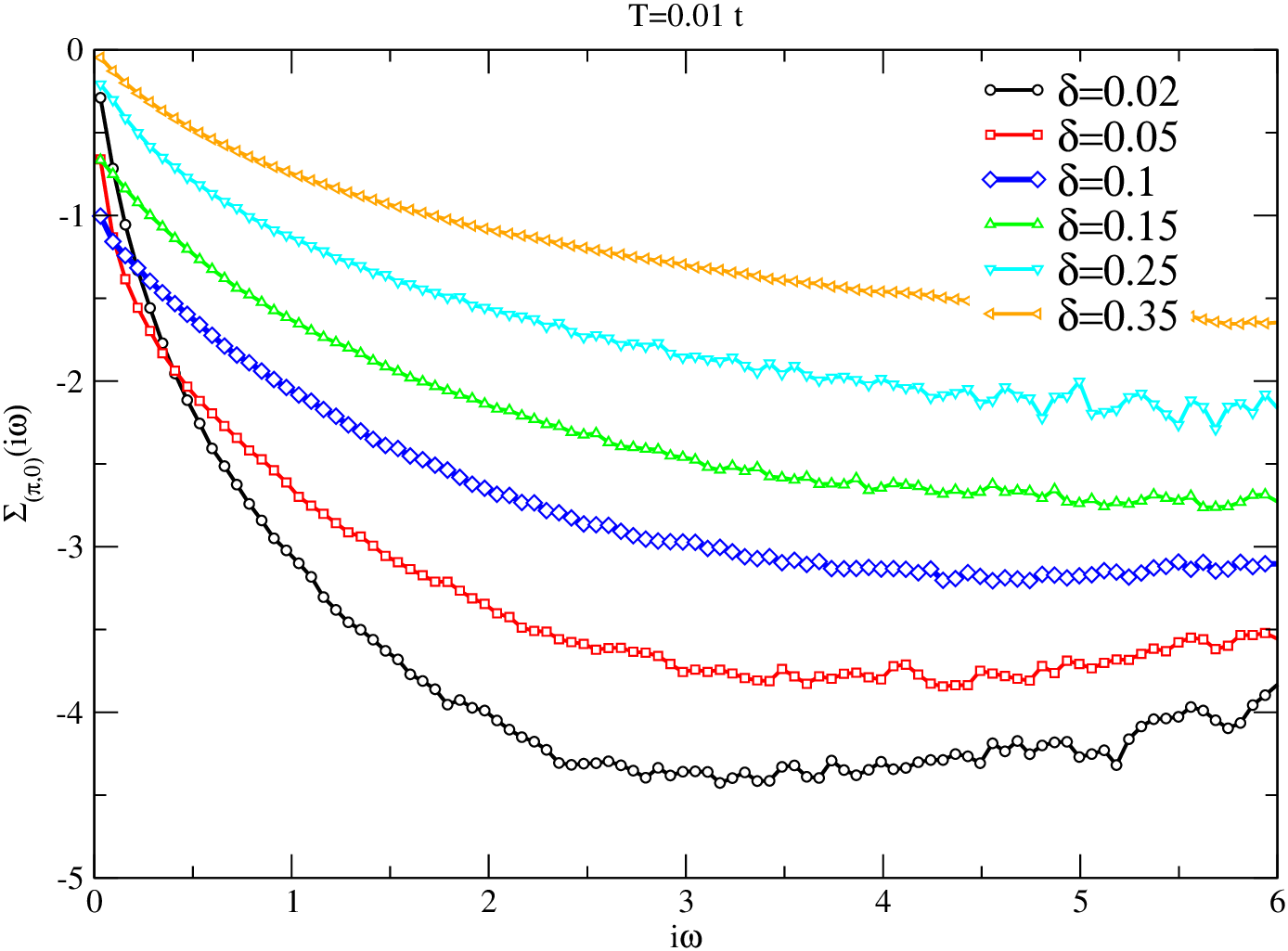}
\caption{ The imaginary part of the cluster self-energy
$\Sigma_{(\pi,0)}$ on imaginary axis. Temperature $T=0.01t$ is around
the critical temperature but still in normal state.  At low frequency,
the most incoherent self-energy corresponds to optimal doped and not
to the underdoped system.  }
\label{Fig2}
\end{figure}

Fig.~\ref{Fig2} shows the imaginary part of the self-energy
$\Sigma_{(\pi,0)}(i\omega)$ at few different dopings and temperature
$T=0.01t$ which is around the superconducting critical temperature of
this approach. The system is still in normal state. It is clear that
the self-energy at large frequencies is a monotonic function of doping
and is largest at the Mott transition $\delta=0$. However, the low
frequency region is distinctly different and the crossing of
self-energies is observed around $\omega_n\sim 0.1 t$. Low doping as
well as large doping self-energies can be extrapolated to zero, while
at optimal doping self-energy remains of the order of unity even around
the critical temperature.

\begin{figure}[htbp]
\includegraphics[width=0.99\linewidth]{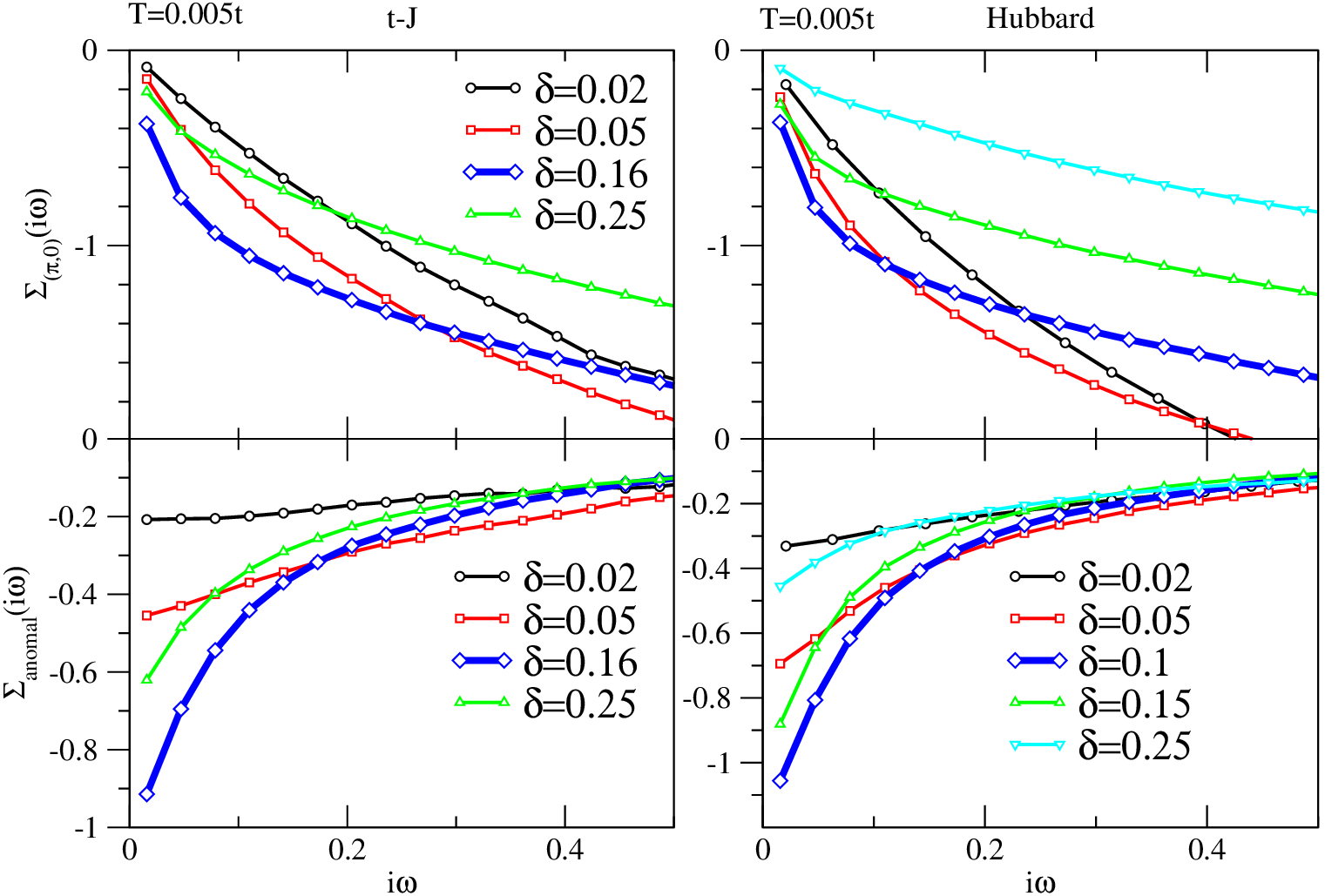}
\caption{
The upper panel shows the cluster $\Sigma_{(\pi,0)}$ self-energy on
the imaginary axis for both the t-J (left) and Hubbard (right) model
showing the large scattering rate at optimal doping. The lower panel shows
the anomalous self-energy for the same models and doping levels and
can be used to locate the optimal doped regime.
}
\label{Fig3}
\end{figure}
The upper panels of Figure \ref{Fig3} show the same self-energy at small
frequency for both the t-J (left) and the Hubbard (right)
model. In this figure, temperature is $T=0.005t$ and is far below $T_c$
therefore all curves extrapolate to zero and the system becomes
coherent in superconducting state. However, the reminiscent of the
strong incoherence at optimal doping is the large slope of imaginary
part of $\Sigma(i\omega)$ which induces very small quasiparticle
residue in this regime.

The precise position of the optimal doping, characterized by the
largest low frequency anomalous self-energy, is different in the t-J
and in the Hubbard model (See lower panel of
Fig.~\ref{Fig3}). It is around $\delta=0.16$ in the former and around
$\delta=0.1$ in the later. However, the large scattering rate is
ultimately connected with the largest anomalous self-energy and hence
largest critical temperature in both the t-J and the Hubbard model.

In our view, the most important advantage of the new Quantum Monte
Carlo method is that it can treat the realistic multiplet structure of
the atom. It was shown in Ref.~\onlinecite{Cm-Pu}  that Hubbard term only ($F_0$),
leads to sever underestimation of the interaction strength in Actinides
and for realistic values of $F_0$, DMFT predicts 
heavy fermion state in Curium rather than magnetic state.
Negligence of the multiplet structure of Plutonium misses the fine
structure of the quasiparticles (two peaks around 0.5 and 0.85 eV) and more
importantly, predicts only weakly correlated metallic state in delta
phase of Plutonium.

\begin{figure}[htbp]
\includegraphics[width=0.99\linewidth]{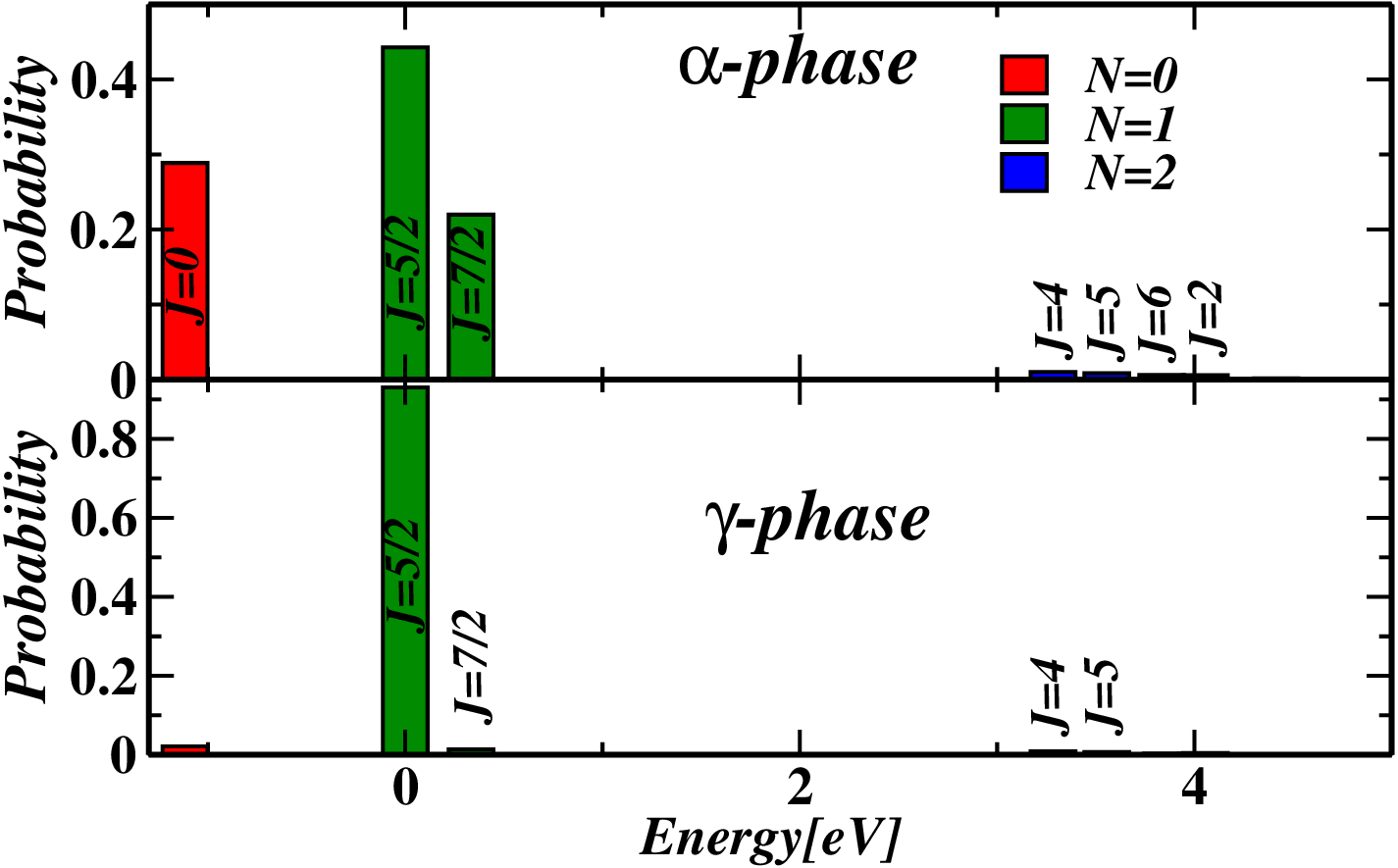}
\caption{ Projection of the DMFT ground state of alpha and gamma
Cerium to various atomic configurations of Cerium atom.  The histograms describe the
generalized concept of valence, where the $f$ electron in the solid
spends appreciable time in a few atomic configurations. The height of
the peak corresponds to the fraction of the time the $f$ electron of
the solid spends in one of the eigenstates of the atom, denoted by the
total spin $J$ of the atom.  We summed up the probabilities for the
atomic states which differ only in the z component of the total spin
$J_z$.  The $x$ axis indicates the energy of atomic eigenstates in the
following way: $Energy(N-1,J)=E_{atom}(N,ground-state)-E_{atom}(N-1,J)$ and
$Energy(N+1,J)= E_{atom}(N+1,J)-E_{atom}(N,ground-state)$, where N is between
0 and 2.}
\label{Fig4}
\end{figure}

To demonstrate the advantage of the new method, we recomputed the
localization-delocalization transition in the archetype material
exhibiting Kondo collapse, namely the $\alpha\rightarrow\gamma$
transition of elemental Cerium. Since the number of electrons in
Cerium fluctuates between states with zero, one and two electrons in
the $f$-shell, the number of atomic states that needs to be kept is
relatively small hence solving the impurity problem requires very little
computational power in this case. In Fig.~\ref{Fig4} we show the
"valence histogram" \cite{Cm-Pu} of the two phases of Cerium, i.e.,
the projection of the density matrix to the eigenstates of the atom.
The plot shows the probability to find an $f$ electron of Cerium in
any of the atomic eigenstates and demonstrates how strongly is the
atom fluctuating between atomic states. The typical fluctuating time
is inversely proportional to the Kondo temperature of the phase, being
around 2000$\,K$ in the alpha phase and around 80$\,$K in the gamma
phase.
The itinerant alpha phase histogram is peaked for many atomic states,
including the spin-orbit split $5/2$ and $7/2$ singly occupied states
as well as the empty state. On the other hand, the local-moment gamma
phase is peaked only at the ground state of the singly occupied sector
with $5/2$ spin showing that the DMFT ground state closely resembles
the atomic $N=1$ ground state.

\section{Conclusion}

We generalized the recently developed continuous time Quantum Monte
Carlo expansion around the atomic limit \cite{Werner_1} to clusters
treated within Cellular DMFT or Dynamical Cluster Approximation as
well as to materials that require realistic Hund's coupling and
multiplet splitting of the atomic state. We explained the steps
necessary for the efficient implementation of the method. Our low
temperature data in the strongly correlated regime of the Hubbard and
t-J model show the efficiency of the method and demonstrate its
superiority compared to the conventional Hirsh-Fye Quantum Monte Carlo
Method. The long-standing problem of adequate treatment of the
multiplet splitting within DMFT is resolved. This splitting is crucial
in actinides and its omission can lead to wrong prediction of the
magnetic nature of the DMFT ground state solution.

We showed that the optimal doped regime in both the t-J and the
Hubbard model is characterized by the largest scattering rate and the
system becomes more coherent in the underdoped and overdoped
regime. The precise position of the optimal doping, as determined from
the maximum in anomalous self-energy, is different in the two
models. However, the strong incoherence is always found at the doping
corresponding to maximum $T_C$.

We computed the valence histogram across the Cerium alpha to gamma
transition with emphasis on the multiplet splitting of the atomic
states. We showed that the atom fluctuates between many atomic states
in itinerant alpha phase and both the $5/2$ and $7/2$ spin-orbit split
states have large probability in the ground state of the system.  The
empty state and the doubly occupied states, which are substantially
split by Hund's coupling, acquire a finite probability. The gamma
phase, on the other hand, shows well defined valence $n_f \sim 1$ and
charge fluctuations become rare.


\section{Acknowledgements}
The author wish to thank G. Kotliar for very stimulating discussion and careful
reading of the manuscript.
 
\appendix

\section{Cellular DMFT in diagonal representation}
\label{AppA}

For the present Continuous Time Monte Carlo method, it is convenient to pick
the base such that the local quantities are diagonal. Cellular DMFT 
uses a generalized open boundary conditions and therefore, in general, momentum is
not a good quantum number. However, for small clusters, such as
plaquette, all sites are equivalent and translational invariance is
still obeyed. The local Green's function and hybridization in momentum
base take the diagonal form
\begin{eqnarray}
  G_{cluster}=\left(
  \begin{array}{cccc}
  G_{(0,0)}&0&0&0\cr
  0&G_{(\pi,0)}&0&0\cr
  0&0&G_{(0,\pi)}&0\cr
  0&0&0&G_{(\pi,\pi)}\cr
  \end{array}
  \right).
\end{eqnarray}
In superconducting state $G_\vK$ is a $2\times 2$ matrix in Nambu
notation. Although the local quantities are diagonal, non-interacting
Hamiltonian at general $\vk$-points is not. The self-consistency
condition in superconducting state takes the following form
\begin{widetext}
\begin{equation}
(i\omega-E_{imp}-\Sigma-\Delta)^{-1}=\sum_\vk\left(
\begin{array}{cc|cc|cc|cc}
  \xi_0(\omega) & -\phi_{0}(i\omega) & -i v_1 & 0 & -i v_2 & 0 & -v_0  & 0\cr
  -\phi_{0}^\dagger(i\omega) & -\xi_0(-i\omega)  & 0 &-i v_1  & 0 & -i
  v_2 & 0 & v_0 \cr
\hline
  i v_1&0&\xi_1(\omega)& -\phi_1(i\omega)& v_0 & 0&-i v_4 & 0\cr
  0& i v_1& -\phi_1^\dagger(i\omega)&  -\xi_1(-i\omega)&  0& -v_0 & 0 & -i v_4   \cr
\hline
 i v_2 & 0 & v_0 & 0 & \xi_2(i\omega) & -\phi_2(i\omega) & -i v_3 & 0 \cr
 0& i v_2 & 0&-v_0& -\phi_2^\dagger(i\omega)&-\xi_2(-i\omega)&0&-i v_3\cr
\hline
-v_0 &0& i v_4 & 0& i v_3 & 0&  \xi_3(\omega)  & -\phi_3(i\omega)\cr
 0 & v_0& 0& i v_4& 0 & i v_3 & -\phi_3^\dagger(i\omega)&-\xi_3(-i\omega)  \cr
\end{array}
\right)
\end{equation}
\end{widetext}
where we defined
\begin{eqnarray}
&& v_0 = t' \sin{k_x}\sin{k_y}\\
&& v_1 = \sin{k_x}(t+t'\cos{k_y})\nonumber\\
&& v_2 = \sin{k_y}(t+t'\cos{k_x})\nonumber\\
&& v_3 = \sin{k_x}(t-t'\cos{k_y})\nonumber\\
&& v_4 = \sin{k_y}(t-t'\cos{k_y})\nonumber\\
&&  \epsilon_0 = -t(2+\cos{k_x}+\cos{k_y})-t'(1+\cos{k_x}\cos{k_y})\nonumber\\
&&  \epsilon_1 = t(\cos{k_x}-\cos{k_y})+t'(1+\cos{k_x}\cos{k_y})\nonumber\\
&&  \epsilon_2 = -t(\cos{k_x}-\cos{k_y})+t'(1+\cos{k_x}\cos{k_y})\nonumber\\
&&  \epsilon_3 =  t(2+\cos{k_x}+\cos{k_y})-t'(1+\cos{k_x}\cos{k_y})\nonumber
\end{eqnarray}
and assumed
\begin{eqnarray}
&& \xi_0 = i\omega+\mu-(\Sigma_{11}+2\Sigma_{12}+\Sigma_{13})-\epsilon_0\\
&& \xi_1 = i\omega+\mu-(\Sigma_{11}-\Sigma_{13})-\epsilon_1\nonumber\\
&& \xi_2 = i\omega+\mu-(\Sigma_{11}-\Sigma_{13})-\epsilon_2 \nonumber\\
&& \xi_3 = i\omega+\mu-(\Sigma_{11}-2\Sigma_{12}+\Sigma_{13})-\epsilon_3\nonumber.
\end{eqnarray}
Here $\Sigma_{11}$ is the normal on-site self-energy, $\Sigma_{12}$ is
the nearest-neighbor and $\Sigma_{13}$ is the next-nearest neighbor
self-energy and $\phi_i$'s are the anomalous components of the
self-energy. For d-wave symmetry, $\phi_0$ and $\phi_3$ vanish and
$\phi_1=-\phi_2$.

The advantage of this formulation of the Cellular DMFT is that the
hybridization becomes block diagonal and hence
the determinants ($Det \underline{\Delta}$) which enter Eq.~(\ref{Z_sample}) can be broken up
into separate contribution for each momentum $\vK$ point like in
Eq.~(\ref{Z1b1}).

%
%


\begin{thebibliography}{99}

\bibitem{Werner_1} P. Werner, A. Comanac, L. De Medici, M. Troyer, A.J. Millis, Phys. Rev. Lett. \textbf{97}, 076405 (2006).

\bibitem{old-review} A. Georges, G. Kotliar, W. Krauth and  M.J. Rozenberg, Rev. Mod. Phys. \textbf{68}, 13 (1996).
\bibitem{phys-today} G. Kotliar, and D. Vollhardt, Physics Today \textbf{57}, 53 (2004).
\bibitem{new-review} G. Kotliar, S.Y. Savrasov, K. Haule, V.S. Oudovenko, O. Parcollet, C.A. Marianetti, Rev. Mod. Phys. {\bf 78}, 865 (2006).

\bibitem{DCA-review} T. Maier, M. Jarrell, T. Pruschke, and M.H. Hettler, Rev. Mod. Phys. \textbf{77}, 1027 (2005).
\bibitem{CDMFT-PRL} G. Kotliar, S.Y. Savrasov, G. Palsson, and G. Biroli, Phys. Rev. Lett. \textbf{87}, 186401 (2001).

\bibitem{Pu-nature} S.Y. Savrasov, G. Kotliar, E. Abrahams, Nature \textbf{410}, 793 (2001).
\bibitem{Am} S.Y. Savrasov, K. Haule, and G. Kotliar, Phys. Rev. Lett. \textbf{96}, 036404 (2006).
\bibitem{Pu-phonons} X. Dai, S.Y. Savrasov, G. Kotliar, A. Migliori, H. Ledbetter, E. Abrahams, Science \textbf{300}, 5621, 953-955 (2003).
\bibitem{Ce-optics} K. Haule, V. Oudovenko, S.Y. Savrasov, and G. Kotliar, Phys. Rev. Lett. \textbf{94}, 036401 (2005).
\bibitem{Cm-Pu} J.H. Shim, K. Haule, and G. Kotliar, cond-mat/0611760.
\bibitem{transport} V.S. Oudovenko, G. Palsson, K. Haule, G. Kotliar,  and S.Y. Savrasov, Phys. Rev. B \textbf{73}, 035120 (2006).


\bibitem{HFQMC} R.M. Fye, and J. E. Hirsch, Phys. Rev. B \textbf{40}, 47804796 (1989);
  J.E. Hirsch, and R. M. Fye, Phys. Rev. Lett. \textbf{56}, 2521 (1986).

\bibitem{ED} M.J. Rozenberg, G. Moeller, and G. Kotliar, Mod. Phys. Lett. B \textbf{8}, 535 (1994);
  M. Caffarel, and W. Krauth, Phys. Rev. Lett. \textbf{72}, 1545 (1994).

\bibitem{NCA} N.E. Bickers, Rev. Mod. Phys. \textbf{59}, 845 (1987).

\bibitem{Pruschke} Th. Pruschke and N. Grewe, Z. Phys. B: Condens. Matter \textbf{74}, 439 (1989).

\bibitem{SUNCA} K. Haule, S. Kirchner, J. Kroha, and P. Wolfle, Phys. Rev. B \textbf{64}, 155111 (2001).

\bibitem{IPT} H. Kajueter and G. Kotliar, Phys. Rev. Lett. \textbf{77}, 131 (1996).
\bibitem{NRG} R. Bulla, Phys. Rev. Lett. \textbf{83}, 136 (1999);
R. Bulla, T. A. Costi, and D. Vollhardt, Phys. Rev. B \textbf{64}, 045103 (2001).
\bibitem{XDai} X. Dai, K. Haule, and G. Kotliar, Phys. Rev. B \textbf{72}, 045111 (2005).

\bibitem{Werner_3} E. Gull, P. Werner, A.J. Millis,  M. Troyer, cond-mat/0609438.

\bibitem{Werner_2} P. Werner, A.J. Millis, Phys. Rev. B \textbf{74}, 155107 (2006).
\bibitem{Werner_1BM} P. Werner, A.J. Millis, cond-mat/0610401.

\bibitem{Chris} C.A. Marianetti, O. Parcollet, K. Haule, G. Kotliar,
  in preparation.
  
\bibitem{Slater-paper} J.C. Slater, Phys. Rev. \textbf{165}, 655 (1968).

\bibitem{Rubtsov} A.N. Rubtsov, V.V. Savkin and A.I. Lichtenstein, Phys. Rev. B \textbf{72}, 035122 (2005).

\bibitem{CTMA} J. Kroha, P. W\"olfle, and T. A. Costi, Phys. Rev. Lett. \textbf{79}, 261 (1997).
  
\bibitem{Anderson} G. Yuval, and P.W. Anderson, Phys. Rev. B
  \textbf{1}, 1522 (1970).

\bibitem{Haldane} F.D.M. Haldane,   Phys. Rev. Lett. \textbf{40}, 416 (1978);
F.D.M.  Haldane, J Phys. C \textbf{11}, 5015 (1978).

\bibitem{Si-Gabi} Q. Si, and G. Kotliar,  Phys. Rev. B 48, 13881 (1993).
  
\bibitem{Cox}
  A. Schiller, F.B. Anders, and D.L. Cox, Phys. Rev. Lett. \textbf{81}, 3235 (1998).
  

\bibitem{Cowan} R.D. Cowan, \textit{The Theory of Atomic Structure and Spectra}, University of California Press, Berkeley, (1981).

\bibitem{QCP-paper} K. Haule, G. Kotliar, cond-mat/0605149.
  
    
\end{thebibliography}
\end{document}